\documentclass[aps,prd,reprint,superscriptaddress,nofootinbib,floatfix]{revtex4-2}
\usepackage{amsmath,amssymb,graphicx,bm,booktabs}
\usepackage[colorlinks=true,allcolors=blue]{hyperref}
\begin{document}
\raggedbottom
\newcommand{\zp}{Z'}
\newcommand{\gev}{\,\mathrm{GeV}}
\newcommand{\mev}{\,\mathrm{MeV}}
\newcommand{\kev}{\,\mathrm{keV}}
\newcommand{\rsun}{R_\odot}

\title{Solar Capture and Suppressed Annihilation of Inelastic Scalar Dark Matter}
\author{XinXin Qi}
\email{qxx@dlut.edu.cn}
\affiliation{Institute of Theoretical Physics, School of Physics, Dalian
University of Technology, No. 2 Linggong Road, Dalian, Liaoning 116024,
P.R. China}
\author{Hao Sun}
\email{haosun@dlut.edu.cn}
\affiliation{Institute of Theoretical Physics, School of Physics, Dalian
University of Technology, No. 2 Linggong Road, Dalian, Liaoning 116024,
P.R. China}
\date{October 2026}

\begin{abstract}
Solar capture of inelastic dark matter does not by itself fix the annihilation signal.
We consider a scalar low-energy model with a $B-3L_\tau$ current, motivated by the high-energy recoil candidate reported by LUX-ZEPLIN.
For $m_\chi=840~\mathrm{GeV}$, a $320~\mathrm{keV}$ splitting, and a $160~\mathrm{MeV}$ mediator, the public LZ response gives 0.965 expected accepted events, while micrOMEGAs gives $\Omega h^2=0.11999834$.
Finite-momentum nuclear responses leave iron as an important solar-capture target despite the near cancellation of its coherent charge at zero momentum transfer.
The captured population remains extended after thermal nuclear motion, elastic and inelastic scattering, and excited-state decay are included, with $r_{\rm eff}=0.215R_\odot$.
Same-state annihilation is open, but the inferred rate is only 0.0404 of the public IceCube $W^+W^-$-calibrated proxy and is smaller than the corresponding isothermal estimate by a factor of 628.
For this benchmark, substantial solar capture is accompanied by a weak neutrino signal.
\end{abstract}

\maketitle

\section{Introduction}
\label{sec:introduction}
The LUX-ZEPLIN (LZ) Collaboration has recently extended its nuclear-recoil
search to energies of about $270\kev$, reporting one event at
$248\pm23\,(\mathrm{stat})\pm23\,(\mathrm{sys})\kev$ in a
$2.84$ tonne-year exposure. For the operators considered by LZ, the largest deviations from the
background-only hypothesis have local and global significances of
$3.4\sigma$ and $2.6\sigma$, respectively~\cite{LZ2026}. The event has
prompted several dark-matter interpretations
\cite{Elahi:2026vlm,Das:2026uyy,Bisal:2026khf,Visinelli:2026kgt,Unwin:2026rdp,
Fan:2026hzw,Lou:2026idn,Bose:2026ndd}. Endothermic scattering is especially
well suited to producing high-energy recoils while suppressing the spectrum
at lower energies. Nearly pure Higgsinos provide a predictive realization
near the thermal mass of $1.1\,\mathrm{TeV}$~\cite{Freese2026,Wu2026},
but are strongly constrained by solar-neutrino searches
\cite{Rodd2026,Pospelov2026}. Xenon nuclear excitation offers a different
possibility, with a characteristic mixed nuclear- and electronic-recoil
signature~\cite{Gu2026}. We instead consider an inelastic transition between
dark-matter internal states, leaving the xenon nucleus in its ground state.

The connection between terrestrial scattering and a solar-neutrino signal
is not one-to-one. Captured dark matter can remain on nonthermal orbits, so
the annihilation rate depends on its subsequent redistribution inside the
Sun. Thermal nuclear motion and departures from capture--annihilation
equilibrium can then be important \cite{Blennow2018}. Recent work motivated
by the LZ event has considered post-capture cooling and thermally assisted
two-state dynamics \cite{Pospelov2026,DiMauro2026}.

Here the dark matter is a split scalar coupled to a light $B-3L_\tau$
mediator. The zero-momentum coherent charge of iron is strongly suppressed,
but the finite-momentum nuclear response restores a sizable capture rate.
After capture we evolve the orbital energy, angular momentum, and internal
state, including thermal nuclear motion, elastic and inelastic scattering,
and excited-state decay. The resulting cloud is spatially extended and has
a small annihilation overlap even though same-state annihilation channels
remain open.

The paper is organized as follows. Section.~\ref{sec:model} introduces the scalar model and the inelastic nuclear transition. Section.~\ref{sec:laboratory} discusses the xenon recoil signal and laboratory constraints. Section.~\ref{sec:capture} presents the solar-capture calculation and the subsequent nonthermal orbital evolution, and Sec.~\ref{sec:neutrinos} derives the corresponding solar-neutrino signal. Section.~\ref{sec:robustness} summarizes the robustness tests, and we give  the discussion and conclusions in the last section. 
\section{Model and nuclear transition}
\label{sec:model}
We extend the split-scalar framework of Ref.~\cite{Qi2022} to a new
low-energy model based on a $B-3L_\tau$ current. All three generations of
quarks carry charge $Q_q=q$, while $L_\tau$ and $\tau_R$ carry $-9q$.
The dark scalar $S$ and the symmetry-breaking field $\Xi$ have charges
$+1$ and $-2$, respectively. An independent exact $Z_2$ symmetry, under
which only $S$ is odd, ensures dark-matter stability. The fermion charge
assignment is anomaly free upon adding a right-handed neutrino with charge
$-9q$; this state and its mass-generation sector are not included in the
low-energy numerical model.

The relevant terms are
\begin{align}
 {\cal L}\supset{}&|D_\mu S|^2+|D_\mu\Xi|^2
 -\frac{\epsilon_{BX}}{2}B_{\mu\nu}X^{\mu\nu}-V,\\
 V\supset{}&M_S^2|S|^2+\lambda_{HS}|H|^2|S|^2
 +\lambda_{\Xi S}|\Xi|^2|S|^2\nonumber\\
 &+\left(\frac{\mu_{S\Xi}}{2}S^2\Xi+\mathrm{H.c.}\right).
\end{align}
We use $\epsilon_{BX}=-\epsilon_{\rm KM}/c_W$, with
$c_W\equiv\cos\theta_W$; Table~\ref{tab:benchmark} lists the positive input
$\epsilon_{\rm KM}$.
The self-quartic is
$\lambda_{DS}|S|^4/4$. After $\Xi=(v_\Xi+\xi)/\sqrt2$ and
$S=(\chi_2+i\chi_1)/\sqrt2$, the last term splits the real states according to
\begin{align}
\label{eq:scalar-splitting}
 m_2^2-m_1^2&=\sqrt2\,\mu_{S\Xi}v_\Xi,\\
 \mu_{S\Xi}&=\frac{\delta(2m_{\chi_1}+\delta)}{\sqrt2v_\Xi}.
\end{align}
The current
\begin{equation}
 J_X^\mu\supset\chi_2\partial^\mu\chi_1-
 \chi_1\partial^\mu\chi_2
\end{equation}
is off diagonal, so one-$\zp$ exchange is endothermic at leading order.
For a nucleus $A$ the minimum speed is
\begin{equation}
 v_{\min}(E_R)=\frac{m_AE_R/\mu_{\chi A}+\delta}
 {\sqrt{2m_AE_R}}.                                      \label{eq:vmin}
\end{equation}
The coherent amplitude is proportional to
$Zf_p+(A-Z)f_n$, with the full propagator
$(m_{\zp}^2+2m_AE_R)^{-1}$ retained.

For each xenon isotope we evaluate the coherent nuclear response
\begin{equation}
\label{eq:nuclear-response}
 W_M(q;f_p,f_n)=c_0^2W_M^{00}+c_1^2W_M^{11}
 +2c_0c_1W_M^{01},
\end{equation}
where $c_0=f_p+f_n$ and $c_1=f_p-f_n$.  Natural isotope abundances are used.
Model conventions and response diagnostics are collected in
Appendixes~\ref{app:model} and \ref{app:xenon}.

\begin{table}[!tb]
\caption{Preferred benchmark and principal observables.  IceCube numbers use
the public spectral proxy and fixed-shape estimate defined in the text. Solar entries refer to the 1024-history Maxwell run shown in Fig.~\ref{fig:solar}.}
\label{tab:benchmark}
\footnotesize
\begin{ruledtabular}
\begin{tabular}{@{}lc@{}}
Quantity & Value\\
\hline
$m_{\chi_1},\ m_{\chi_2}-m_{\chi_1}$ & $840\gev,\ 320\kev$\\
$m_{\zp},\ m_{h_2}$ & $160\mev,\ 30\gev$\\
$g_X,\ Q_\chi,\ Q_q$ & $3.1692\times10^{-3},\ 1,\ -2.2222\times10^{-3}$\\
$\epsilon_{KM},\ \lambda_{\Xi S}$ & $1.4489\times10^{-4},\ 0.520346$\\
$\sin\alpha,\ \lambda_{HS}$ & $10^{-5},\ -1.49\times10^{-7}$\\
$\lambda_{DS},\ \mu_{S\Xi}$ & $0.1,\ 15.0592\mev$\\
$f_n/f_p$ & $-0.869850$\\
$\Omega h^2$ & $0.11999834$\\
LZ expected yield, $125$--$400\kev$ & $0.964605$\\
$C_\odot$ & $9.12956\times10^{19}\,{\rm s}^{-1}$\\
$r_{\rm eff}/\rsun$ & $0.2153$\\
$\Gamma_{\rm shape}$ & $7.2535\times10^{16}\,{\rm s}^{-1}$\\
$\Gamma_{\rm shape}/\Gamma_{\rm IC}^{90}$ & $0.03676$--$0.04037$\\
\end{tabular}
\end{ruledtabular}
\end{table}

\section{Xenon signal and laboratory constraints}
\label{sec:laboratory}
\subsection{Xenon spectrum, relic abundance, and NA64 constraints}
For the benchmark point, diagonalizing the neutral-vector sector gives
$g_\chi=3.16918\times10^{-3}$,
$f_p=2.42891\times10^{-5}$, and
$f_n=-2.11279\times10^{-5}$. The xenon recoil spectrum uses natural isotope
abundances and the isotope-dependent coherent $W_M$ response.

\begin{figure}[tb]
\includegraphics[width=\columnwidth]{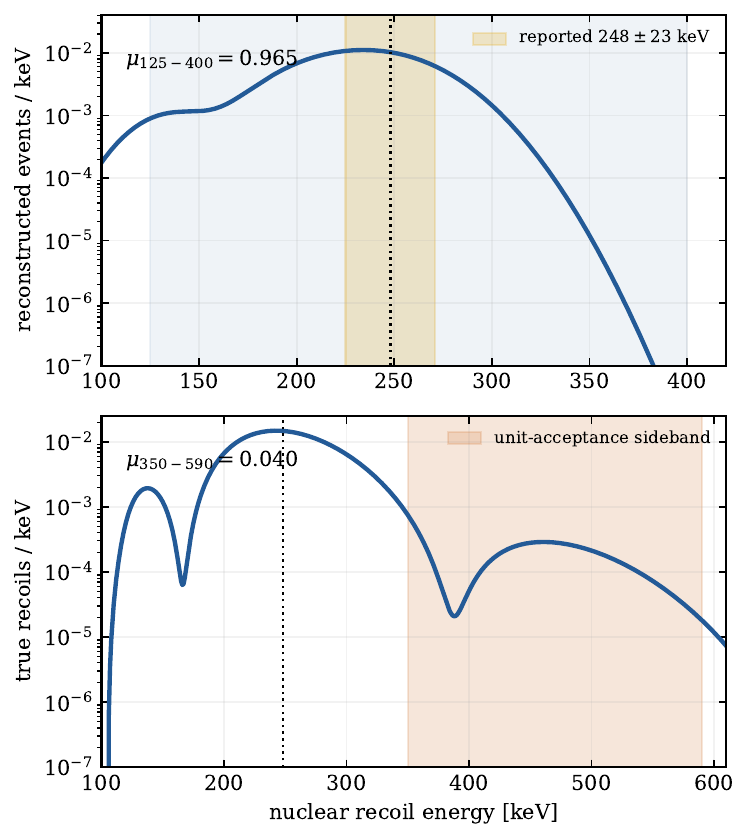}
\caption{Current $B-3L_\tau$ benchmark spectrum.  Top: reconstructed spectrum
using the public final-region LZ efficiency.  Bottom: true recoil spectrum;
unit acceptance in the indicated sideband makes its integral conservative.}
\label{fig:lz}
\end{figure}

For Fig.~\ref{fig:lz}, the recoil spectrum is folded with the published LZ
nuclear-recoil efficiency and a Gaussian resolution,
$\sigma_E=1.46\sqrt{E_R/\mathrm{keV}}\kev$. The reconstructed
$125$--$400\,\mathrm{keV}$ interval contains $0.964605$ events, while the
true-recoil spectrum contains only $0.0402644$ events between $350$ and
$590\,\mathrm{keV}$ even for unit acceptance. The calculation uses the public LZ response; the full experimental likelihood is not available.

We implement the model in FeynRules and CalcHEP and evaluate the relic
density with micrOMEGAs~\cite{FeynRules,CalcHEP,micromegas}. Freeze-out is
split almost equally between $\chi_i\chi_i\to Z^{\prime}Z^{\prime}$ and
$\chi_i\chi_i\to h_2h_2$. At the present epoch,
$\langle\sigma v\rangle_0=4.45787\times10^{-26}\,\mathrm{cm^3\,s^{-1}}$.
The $Z^{\prime}$ branching fractions are
$\mathrm{Br}(Z^{\prime}\to e^+e^-)=0.5066$ and
$\mathrm{Br}(Z^{\prime}\to\nu_\tau\bar{\nu}_\tau)=0.4934$.

For the NA64 recast we use the published scaling
$N_{\rm sig}\propto(g_{eV}^2+g_{eA}^2)\mathrm{Br}_{\rm inv}$ for
bremsstrahlung and resonant production~\cite{NA642026}. At
$m_{Z'}=160~\mathrm{MeV}<2m_\mu$ we find
$g_{eV}=-4.54170\times10^{-5}$ and $|g_{eA}|=4.56\times10^{-11}$.
The canonical unbroken $B-L$ model has $\mathrm{Br}_{\rm inv}=0.75$ at this
mass, so the benchmark corresponds to
\begin{equation}
g_{B-L}^{\rm eff}
=
\sqrt{
(g_{eV}^2+g_{eA}^2)
\frac{0.4933735}{0.75}
}
=
3.68363\times10^{-5},
\end{equation}
which is below the observed 90\% C.L. limit. The same point also satisfies
the $B-3L_\tau$ oscillation bound quoted in Ref.~\cite{HeeckNSI2019}:
$m_{Z'}/|g_{B-3L_\tau}|=7.57~\mathrm{TeV}$, compared with the
$4.8~\mathrm{TeV}$ reference value. Appendix~\ref{app:na64} gives the
NA64 recast details.

\section{Capture and nonthermal solar evolution}
\label{sec:capture}
\subsection{Thermal capture}
In the stationary-target limit, for incident speed $u$ and local speed
$w^2=u^2+v_{\rm esc}^2(r)$, the inelastic transition is open when
$w^2>2\delta/\mu_A$.  The allowed recoil endpoints and the capture threshold
are
\begin{align}
E_R^\pm&=\frac{\mu_A^2}{2m_A}
\left(w\pm\sqrt{w^2-\frac{2\delta}{\mu_A}}\right)^2,\\
E_R^{\min}&=\max\left(E_R^-,\frac{m_\chi u^2}{2}-\delta,0\right).
\end{align}

For moving nuclei, we sample their thermal velocities and require the outgoing
dark matter to remain gravitationally bound in the solar frame.
For normalized incident speed distribution $F(u)$, the thermal capture rate is
\begin{equation}
 C_\odot=\frac{\rho_\chi}{m_\chi}\int4\pi r^2dr
 \int du\,\frac{F(u)}u\,w\,\Omega_{\rm cap}(w,r),
 \label{eq:capture}
\end{equation}
where $w^2=u^2+v_{\rm esc}^2(r)$, and $\Omega_{\rm cap}$ includes the
sum over solar isotopes and the thermal average over nuclear velocities,
requiring the scattered dark matter to remain bound. We use
$\rho_\chi=0.4\,\mathrm{GeV\,cm^{-3}}$, $v_0=238\,\mathrm{km\,s^{-1}}$,
$v_\odot=250.591\,\mathrm{km\,s^{-1}}$, and
$v_{\rm esc}^{\rm gal}=544\,\mathrm{km\,s^{-1}}$. The same halo model is
used for LZ, including the time-averaged motion of the Earth.

We use the micrOMEGAs solar composition with 42 isotopes, retaining the
full mediator propagator and the available shell-model $W_M$ responses,
with Helm form factors used otherwise~\cite{Catena2015}. We obtain
$C_\odot=9.12956\times10^{19}\,\mathrm{s^{-1}}$, of which $^{56}$Fe
accounts for $64.7\%$. Despite the near cancellation
$26+30f_n/f_p\simeq-0.10$ at zero momentum transfer, iron remains a
major capture target. 

\subsection{Finite-rate orbital evolution}
Each captured particle is drawn from the differential capture distribution
and subsequently evolved in the solar potential. For a spherically symmetric
Sun, we characterize an orbit by its specific energy and angular momentum,
$(E,L)$. The radial turning points are determined by
\begin{equation}
v_r^2(r)=2[E-\Phi(r)]-\frac{L^2}{r^2}=0,
\end{equation}
with the time spent along the orbit weighted by $dr/|v_r|$. For each solar
isotope, we average the elastic and state-changing scattering rates,
$n_A(r)\langle\sigma_A v_{\rm rel}\rangle$, over the orbit.

The time to the next interaction is drawn from the total scattering rate.
Once an interaction occurs, its position along the orbit, the nuclear
thermal velocity, recoil kinematics, and outgoing direction are generated
from the corresponding distributions. The evolution continues until the
assigned particle age is reached or the particle escapes the Sun. To
represent continuous capture over the solar lifetime, particle ages are
drawn uniformly from $0$ to $4.57\,\mathrm{Gyr}$. Annihilation is not
included during the trajectory evolution and is evaluated separately from
the resulting spatial distribution.
Both dark states are followed, together with elastic and inelastic
scattering, excited-state decay, and escape. The transport sample shown below
was generated with $\tau_2=136.6$ days; the updated benchmark gives
$\tau_2=137.2$ days. Its scalar elastic nucleon cross section is $0.108$ of
the value used in that sample. The corresponding lifetime and elastic-rate
tests are given in Appendix~\ref{app:systematics}, and we retain the same
transport sample for the quoted central orbital quantities. Appendix~\ref{app:capture}
gives the decay-width and sampling details.

\subsection{Residence distribution and annihilation overlap}
\begin{figure}[tb]
\centering
\includegraphics[width=0.90\columnwidth]{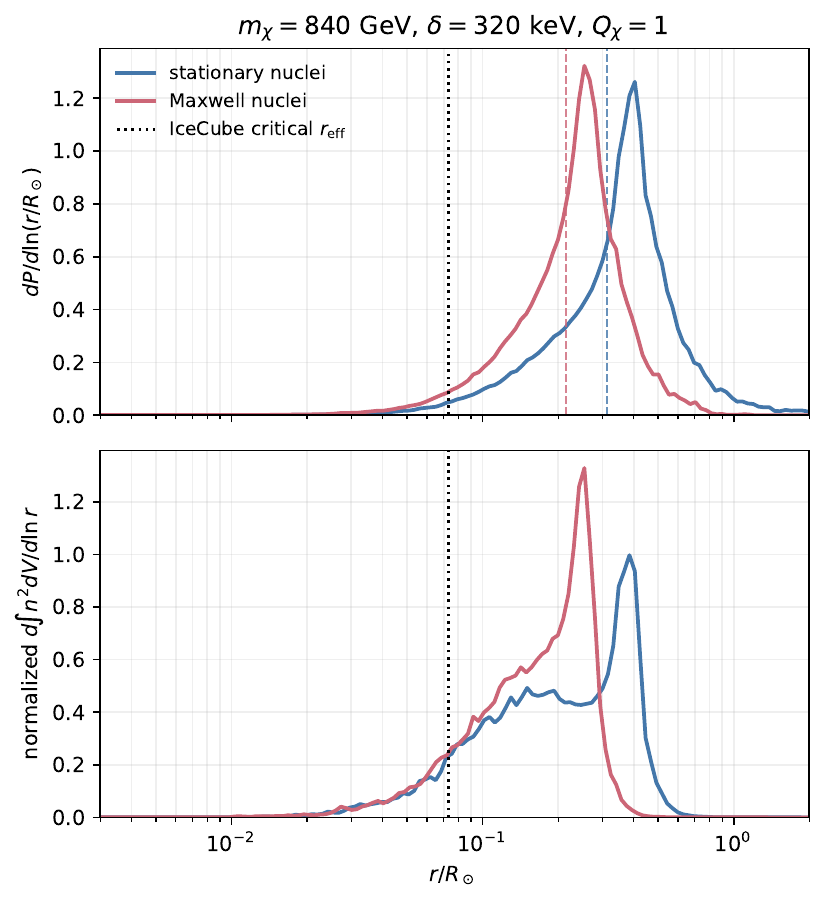}
\caption{Orbit-residence distribution (top) and annihilation overlap
(bottom).  Thermal nuclear motion contracts the cloud relative to the
stationary-nucleus control. Each curve contains 1024 capture histories with
scalar elastic, inelastic and decay processes. Colored dashed lines mark
$r_{\rm eff}$; the black dotted line uses the Maxwell retained-capture
normalization and the fixed-shape $WW$ proxy. It is a diagnostic reference,
not an official exclusion radius.}
\label{fig:solar}
\end{figure}

For a normalized residence density $p(r)$, we define
\begin{equation}
 \begin{gathered}
 \int p\,dV=1,\qquad
 V_{\rm eff}^{-1}=\int p^2\,dV,\\
 r_{\rm eff}=\left[\frac{V_{\rm eff}}{(2\pi)^{3/2}}\right]^{1/3}.
 \end{gathered}
\end{equation}
For Maxwellian nuclear motion,
$r_{\rm eff}=0.2153\rsun$; with stationary nuclei it is $0.3141\rsun$.
The capture calculation and the orbital evolution in Fig.~\ref{fig:solar}
use the same treatment of nuclear motion.

\section{Solar-neutrino signal}
\label{sec:neutrinos}
For the present-day annihilation, micrOMEGAs finds nearly equal
contributions from $\chi_1\chi_1\to\zp\zp$ and
$\chi_1\chi_1\to h_2h_2$, with fractions $0.500076$ and $0.499924$,
respectively. With $\mathrm{Br}(h_2\to\zp\zp)\simeq1$ and
$\mathrm{Br}(\zp\to\nu_\tau\bar\nu_\tau)=0.4933735$, the resulting
cascades produce an average of $2.960$ prompt neutrinos and antineutrinos
per annihilation.

We propagate the box-shaped neutrino spectra through the Sun using the
PPPC4DMnu tables implemented in micrOMEGAs and fold the result with the
public IceCube \cite{Catena:2018vzc} effective areas. Normalizing to the published 2025
$\nu_\tau\bar\nu_\tau$ and $W^+W^-$ limits at $840\gev$ gives
\begin{align}
\Gamma_{\rm IC}^{90}(\nu_\tau\ {\rm calibration})
&=1.9730\times10^{18}\,{\rm s^{-1}},\\
\Gamma_{\rm IC}^{90}(WW\ {\rm calibration})
&=1.7968\times10^{18}\,{\rm s^{-1}}.
\end{align}
The two calibrations differ by $9.8\%$. For the radial profile above, the
annihilation rate is estimated from
\begin{equation}
 \Gamma_{\rm shape}
 =\frac{C_b}{2}\tanh^2\!\left[
 T\sqrt{\frac{C_b\langle\sigma v\rangle}{V_{\rm eff}}}
 \right],
\end{equation}
where $C_b$ is the retained capture rate. The fixed-shape reference radius
is set by $\Gamma_{\rm shape}=\Gamma_{\rm IC,WW}^{90}$. Figure~\ref{fig:solar}
uses the Maxwellian normalization, whereas the mediator-mass scan uses the
largest reference radius in the sampled set. For the Maxwell orbit sample,
$\Gamma_{\rm shape}=7.25\times10^{16}\,\mathrm{s^{-1}}$, corresponding to
$0.037$--$0.040$ of the IceCube calibration. The calculation uses public
effective areas and channel limits rather than an event-level likelihood;
the reference radius is only a diagnostic quantity. No injection-radius
dependence is included in the spectral calibration. The overlap and
compact-core checks are listed in Appendix~\ref{app:icecube}.

\subsection{Comparison with an isothermal distribution}
For comparison with a thermalized core, we keep the retained capture rate,
annihilation cross section, and $WW$ calibration fixed at their benchmark
values,
\begin{align*}
C_b&=9.11172663\times10^{19}\,\mathrm{s}^{-1},\\
\langle\sigma v\rangle&=4.45787169\times10^{-26}\,
\mathrm{cm^3\,s^{-1}} .
\end{align*}
An isothermal distribution in the same solar potential, evaluated at
$T_c=1.567\times10^7\,\mathrm{K}$, gives
$r_{\rm eff}=0.00366607R_\odot$ and
$V_{\rm eff}=2.61638\times10^{26}\,\mathrm{cm^3}$. The corresponding
fixed-shape estimate has an equilibration time of
$2.5433\times10^8$ yr and
$\Gamma_{\rm shape}=4.55586\times10^{19}\,\mathrm{s}^{-1}$,
or $25.355$ times the $WW$ proxy. For the sampled benchmark orbits, the
same ratio is only $0.0403652$, giving a difference of a factor of
$628.14$ in the annihilation rate.

No thermalized population is evolved and its escape rate is not recomputed;
both cases use the same neutrino spectral calibration.

\section{Robustness}
\label{sec:robustness}
The mediator-mass transport scan gives
$r_{\rm eff}/\rsun=0.2077$--$0.2239$ and fixed-shape $WW$ ratios of
$0.023$--$0.066$. With the sensitivity variations included, the ratio spans
$0.015$--$0.137$, below the proxy limit throughout.

Varying either the excited-state lifetime or the elastic cross section by a
factor of ten gives $r_{\rm eff}/\rsun=0.2107$--$0.2147$ and fixed-shape
$WW$ ratios of $0.0408$--$0.0431$. They are transport tests rather than additional
self-consistent model points. Appendix~\ref{app:systematics} contains the
full set of transport and solar-profile runs.

Using the SF-III/GS98 and SF-III/AGSS09 solar profiles~\cite{ICE2023}
reduces the capture rate to about $0.74$ and $0.68$ of the baseline value.
The corresponding orbit calculations give $r_{\rm eff}/\rsun=0.217$--$0.222$
and fixed-shape $WW$ ratios of $0.017$--$0.033$ (Appendix~\ref{app:systematics}),
again below the public IceCube proxy.

\section{Discussion and conclusions}
\label{sec:discussion}
The benchmark passes the tree-level boundedness and numerical vacuum checks.
Its elastic proton cross section is
$\sigma_p^{\rm SI}=9.16\times10^{-55}\,\mathrm{cm^2}$. The micrOMEGAs
cascade spectrum gives a Planck ratio of $0.02571$, and the approximate
four-dwarf limits are also satisfied. Relic, decay, and indirect-detection
calculations use the low-energy model defined above; no ultraviolet completion
is assumed. The solar transport sample is the one described in
Sec.~\ref{sec:capture}. Further checks are collected in
Appendix~\ref{app:systematics}.

Finite-momentum nuclear effects leave iron as an efficient capture target,
but the captured population does not collapse to a compact core. At fixed
capture rate, annihilation cross section, and spectral calibration, the
isothermal comparison in Sec.~\ref{sec:neutrinos} changes the annihilation
rate by more than two orders of magnitude. Since
$\chi_1\chi_1\to Z'Z'$ and $\chi_1\chi_1\to h_2h_2$ are both open, the
suppression comes from the spatial distribution rather than from a shortage
of annihilating partners \cite{DiMauro2026}.

Every transport case considered lies below the public IceCube spectral
proxy. The ratio is sensitive to the nuclear response, rare compact orbits,
higher-order elastic effects, and the use of public detector information; an
unresolved isothermal core would raise it to about $0.23$. In all of these
cases the captured population remains too extended to produce an annihilation
rate comparable to the IceCube proxy.

\begin{acknowledgments}
Hao Sun is supported by the National Natural Science Foundation of China
(Grant Nos.\ 12075043 and 12147205). ChatGPT 6.0 (Astra) was used for code
debugging, numerical cross-checks, literature comparison, and language
editing. The authors checked the code changes and numerical results and take
responsibility for the scientific conclusions.

\end{acknowledgments}

\appendix
\section{Model conventions and scalar-spectrum checks}
\label{app:model}
\subsection{Gauge charges and anomaly cancellation}
The gauge generator is proportional to $B-3L_\tau$; its overall sign and
normalization are free.  The normalization used for the benchmark is
\begin{center}
\begin{tabular}{lcc}
\hline\hline
Field & $Q_X$ & Benchmark\\
\hline
$Q_L,u_R,d_R$ & $q$ & $-2.22222\times10^{-3}$\\
$L_\tau$ & $-9q$ & $2.00000\times10^{-2}$\\
$\tau_R$ & $-9q$ & $2.00000\times10^{-2}$\\
$N_{\tau R}$ & $-9q$ & $2.00000\times10^{-2}$\\
$S$ & $1$ & $1$\\
$\Xi$ & $-2$ & $-2$\\
\hline\hline
\end{tabular}
\end{center}
All other Standard Model leptons are neutral.  An exact $Z_2$ is imposed,
under which $S$ is odd and every other field is even, as in the scalar model
of Ref.~\cite{Qi2022}.  Dark-matter stability does not rely on a particular remnant of $U(1)_X$
breaking.

Using left-handed Weyl fields, the mixed non-Abelian anomalies are
\begin{align}
 {\cal A}_{SU(3)^2X}&=3(q-q/2-q/2)=0,\\
 {\cal A}_{SU(2)^2X}&=(9q-9q)/2=0.
\end{align}
The sums for $Y^2X$, $YX^2$, $X^3$, and gravity$^2X$ also vanish after one
$N_{\tau R}$ is included.  

The field $N_{\tau R}$ is included only for anomaly cancellation and is not
propagated in the numerical model. The calculation contains the Standard Model
fields, $S$, $\Xi$, and the new vector boson. No ultraviolet completion is
specified.

\subsection{Scalar spectrum}
The scalar convention is
\begin{align}
V\supset{}&M_S^2|S|^2+\lambda_{HS}|H|^2|S|^2
+\lambda_{\Xi S}|\Xi|^2|S|^2+\frac{\lambda_{DS}}4|S|^4\nonumber\\
&+\left(\frac{\mu_{S\Xi}}2S^2\Xi+\mathrm{H.c.}\right).
\end{align}
The benchmark uses $\sin\alpha=10^{-5}$, $\lambda_{HS}=-1.49\times10^{-7}$,
$\lambda_{\Xi S}=0.5203460264$, $\lambda_{DS}=0.1$, and bare
$M_S=839.90147445\gev$.

Using Eq.~\eqref{eq:scalar-splitting}, for $v_\Xi=25.24314717\gev$ the benchmark uses
$\mu_{S\Xi}=15.0591634\,\mathrm{MeV}$.  The CalcHEP spectrum is
$m_{\chi_1}=840.0000000\gev$ and
$m_{\chi_2}-m_{\chi_1}=320.0000\kev$.

\section{Xenon-response and mass--splitting diagnostics}
\label{app:xenon}
\subsection{Physical couplings and detector response}
We diagonalize hypercharge-$X$ kinetic mixing and the neutral-vector mass
matrix before extracting couplings. We write the kinetic term as
$-\epsilon_{BX}B_{\mu\nu}X^{\mu\nu}/2$, with
$\epsilon_{BX}=-\epsilon_{\rm KM}/c_W$ and
$\epsilon_{\rm KM}=1.44893353\times10^{-4}$.  At the preferred point,
\begin{align}
g_\chi&=3.16917693\times10^{-3},\\
g_V^u&=2.32353785\times10^{-5},\\
g_V^d&=-2.21816370\times10^{-5},\\
f_p&=2g_V^u+g_V^d=2.42891201\times10^{-5},\\
f_n&=g_V^u+2g_V^d=-2.11278954\times10^{-5}.
\end{align}
The largest quark axial coefficient divided by the largest vector coefficient
is $1.96\times10^{-6}$.

Using the isotope-dependent response of Eq.~\eqref{eq:nuclear-response},
after the public final-region efficiency, the leading fractions are Xe-132
($26.1\%$), Xe-136 ($20.4\%$), Xe-131 ($19.2\%$), Xe-134 ($16.0\%$), and
Xe-129 ($15.0\%$).

The true rate is folded with
\begin{align}
G(E_{\rm rec},E_R)&=\frac{
e^{-(E_{\rm rec}-E_R)^2/(2\sigma_E^2)}}{\sqrt{2\pi}\sigma_E},\\
\sigma_E&=1.46\sqrt{E_R/\mathrm{keV}}\kev.
\end{align}
The central final-region efficiency is taken from the vector data for the
official LZ Fig.~S2~\cite{LZ2026}. For an exposure of
$1.03731\times10^6\,\mathrm{kg\,day}$, the reconstructed spectrum contains
$0.964605$ events in $125$--$400\kev$ and peaks near $234\kev$. With unit
acceptance, the true-recoil spectrum contains $0.0402644$ events between
$350$ and $590\kev$. Because no probability model is provided for the
efficiency envelope, we use its central curve and do not profile it as a
nuisance parameter. We use it as a public-response reconstruction, not as the official LZ
likelihood.

\subsection{Mass--splitting projection}
We vary $m_{\chi_1}=550$--$1300\gev$ and
$\delta=275$--$360\kev$, keeping the benchmark couplings and
$m_{\zp}$ fixed. At each point we recalculate the xenon recoil spectrum
using all nine natural isotopes, the 48 time bins, the finite mediator
propagator, the $W_M$ response, and the LZ efficiency and energy
resolution. Within the public background proxy, the benchmark lies close to the best
point on this fixed-coupling grid, with $-2\Delta\ln{\cal L}=0.01775$.
The descriptive $q<2.30$ band spans $\delta=290$--$357.5\kev$ over the
mass range shown. The band is used only as a mass--splitting diagnostic and
is not a statistical confidence contour.

\begin{figure*}[t]
\includegraphics[width=0.90\textwidth]{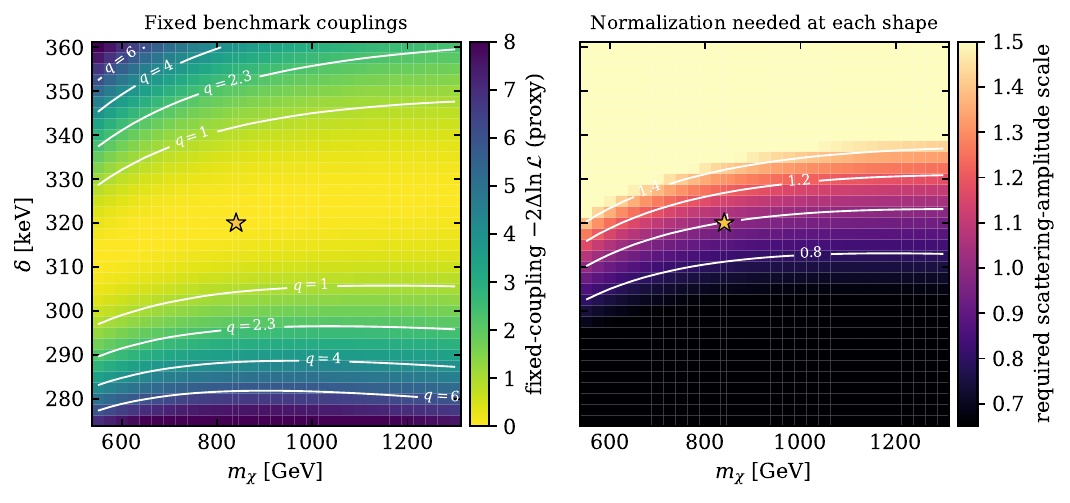}
\caption{Public LZ mass-splitting scan.  Left: relative likelihood on the
fixed-benchmark-coupling slice.  Right: scattering-amplitude rescaling selected
when the normalization is profiled independently at each spectral shape.  The
contours are public-response diagnostics and must not be interpreted as
official LZ confidence regions.}
\label{fig:massdelta}
\end{figure*}

\section{Solar-capture and orbital-transport validation}
\label{app:capture}
\subsection{Capture integration and nuclear-response controls}
The capture calculation follows the thermal treatment described in
Sec.~\ref{sec:capture}. The halo parameters are taken to be
$\rho_\chi=0.4\,\mathrm{GeV\,cm^{-3}}$,
$v_0=238\,\mathrm{km\,s^{-1}}$,
$v_\odot=250.591\,\mathrm{km\,s^{-1}}$, and
$v_{\rm esc}^{\rm gal}=544\,\mathrm{km\,s^{-1}}$,
matching those used in the terrestrial analysis.

The micrOMEGAs solar profile includes 42 isotopes. Shell-model
$W_M$ responses~\cite{Catena2015} are available for isotopes accounting
for about $91.7\%$ of the initial capture rate; for the remaining isotopes
we use Helm form factors. The full mediator propagator is retained
throughout. For the Maxwellian benchmark we obtain
$C_\odot=9.1295578\times10^{19}\,\mathrm{s^{-1}}$, with $^{56}$Fe
contributing $64.7\%$. The near cancellation of the iron charge at zero momentum transfer does not
remove its finite-momentum contribution. Replacing the shell-model responses by common proton and neutron
Helm form factors lowers the total capture rate by a factor of $3.18$ and
the $^{56}$Fe contribution by a factor of $197$.

On a reduced grid at $m_{Z'}=160~\mathrm{MeV}$ we obtain
$C_\odot=9.1333038\times10^{19}\,\mathrm{s^{-1}}$, $0.041\%$ above the
benchmark result. Independent scrambled-Sobol integrations for the baseline
and alternative solar profiles differ by at most $0.194\%$. We use this
spread as a numerical-integration check only.

\paragraph{Matched nuclear-response control.}
We repeat the thermal capture calculation with the benchmark parameters fixed
and change only the nuclear form-factor prescription. Common proton and
neutron Helm form factors give
$C_\odot=2.86826\times10^{19}\,\mathrm{s^{-1}}$ and
$C_{^{56}\mathrm{Fe}}=2.99263\times10^{17}\,\mathrm{s^{-1}}$.
Using the available shell-model $W_M$ responses, with Helm form factors for
the remaining isotopes, gives $9.11925\times10^{19}$ and
$5.90674\times10^{19}\,\mathrm{s^{-1}}$, respectively. Thermal nuclear
motion and the full mediator propagator are retained in both calculations.

Each value is the average of two scrambled-Sobol integrations with $2^{18}$
points. The realizations differ by $0.094\%$ for common Helm form factors and
by $0.028\%$ for the shell-model calculation. Common Helm form factors retain
the finite-momentum suppression but also retain the factorized proton--neutron
cancellation inherited from zero momentum transfer. The main-text capture
rate comes from the higher-statistics shell-model run.

\subsection{Decay width and orbital sampling}
Excited-state decay competes with down-scattering. A full
CalcHEP/micrOMEGAs three-body integration gives
\begin{equation}
\Gamma(\chi_2\to\chi_1\nu_\tau\bar\nu_\tau)
=5.55247\times10^{-32}\gev,
\end{equation}
or $\tau_2=137.20$ days.  The small-splitting contact expression
\begin{equation}
\Gamma=\frac{g_\chi^2g_{L\nu_\tau}^2\delta^5}
{120\pi^3m_{\zp}^4}
\end{equation}
differs little from the full three-body result. The transport calculation
uses the full width.

\begin{table*}[t]
\caption{The stationary and Maxwell benchmark histories used in Fig.~\ref{fig:solar}.
Both include tree-level scalar elastic scattering. $\Gamma_{\rm shape}$ is
the fixed-shape diagnostic; the ratio range uses the two spectral calibrations.}
\label{tab:orbit}\small
\begin{ruledtabular}\begin{tabular}{lcc}
 & stationary nuclei & Maxwell nuclei\\\hline
histories & 1024 & 1024\\
bound histories & 1021 & 1022\\
median $r_{\max}/\rsun$ & 0.46099 & 0.30051\\
$r_{\rm eff}/\rsun$ & 0.31406 & 0.21532\\
$\Gamma_{\rm shape}$ [s$^{-1}$] & $2.04713\times10^{16}$ & $7.25346\times10^{16}$\\
$\Gamma_{\rm shape}/\Gamma_{\rm IC}^{90}$ & 0.01038--0.01139 & 0.03676--0.04037\\
\end{tabular}\end{ruledtabular}\end{table*}

The annihilation overlap is formed from pairs of independent capture
histories; self-pairs are omitted. For radial-bin probabilities $p_{ib}$, the
pair contribution is divided by the corresponding bin volume. Escaped
histories carry zero weight in the captured-population normalization.
Bootstrap resampling gives 16--84\% ranges of $0.31087$--$0.31702\rsun$
for stationary nuclei and $0.21295$--$0.21697\rsun$ for Maxwellian nuclei.
No retained orbit in either sample has an apocentre below $0.01\rsun$.

\section{Annihilation-overlap diagnostics}
\label{app:icecube}
A second estimate uses the no-annihilation-feedback orbital distribution,
\begin{equation}
\widehat\Gamma_0=
\frac{\langle\sigma v\rangle(CT)^2}{2N(N-1)}
\sum_{i\ne j,b}\frac{p_{ib}p_{jb}}{V_b}.
\end{equation}
All initial histories enter the normalization, with escaped trajectories
assigned zero weight. In the linear transport problem an annihilation sink can
only lower the exact rate relative to the no-sink solution, although a finite
Monte Carlo sample need not preserve this ordering exactly.

The $WW$ reference radius in Fig.~\ref{fig:solar} is $0.07326\rsun$.
With 768 orbit nodes and 600 radial bins,
$\widehat\Gamma_0/\Gamma_{\rm IC,WW}^{90}=0.04045$; the elastic-stress run
gives $0.04420$. Adding an unresolved isothermal core raises these values to
$0.2229$ and $0.2316$. They are sensitivity estimates, not confidence limits.

\section{NA64 recast validation}
\label{app:na64}
At $m_{Z'}=160\,\mathrm{MeV}<2m_\mu$, the benchmark electron couplings are
$g_{eV}=-4.54170\times10^{-5}$ and
$|g_{eA}|=4.56\times10^{-11}$. The 2026 NA64 analysis gives the invisible
signal scaling
\begin{equation}
N_{\rm sig}\propto
(g_{eV}^2+g_{eA}^2)\,\mathrm{Br}_{\rm inv},
\end{equation}
for radiative production, with the resonant contribution scaling
equivalently as
$\Gamma_{e^+e^-}\mathrm{Br}_{\rm inv}$~\cite{NA642026}.

The vector data for NA64 Fig.~6 give
$g_{B-L}^{90}=1.31432\times10^{-4}$. The benchmark yield is $7.86\%$ of the
limit, or $8.18\%$ after shifting the digitized curve downward by $2\%$.
No event-level detector likelihood enters this recast.

\section{Robustness and theoretical consistency}
\label{app:systematics}
\subsection{Solar and transport sensitivity}
The mediator-mass scan is used as a transport test rather than as the final
laboratory-allowed region. At fixed $m_{Z'}$, the LZ normalization fixes
$g_\chi f_p$ and $g_\chi f_n$, so a single initial capture sample can be used
along each mass slice. Elastic amplitudes and relic-density parameters are
recalculated point by point.

The capture rate varies from
$7.34642\times10^{19}\,\mathrm{s^{-1}}$ at $m_{Z'}=200\mev$ to
$1.12072\times10^{20}\,\mathrm{s^{-1}}$ at $m_{Z'}=130\mev$.
For the representative points used in the orbit calculation, we find
\begin{align}
r_{\rm eff}/\rsun &= 0.20774\text{--}0.22393,\\
\Gamma_{\rm shape}/\Gamma_{\rm IC,WW}^{90}
&=0.02333\text{--}0.06626.
\end{align}
The largest fixed-shape ratio occurs near
$m_{Z'}=130\mev$ and $g_X=1.04\times10^{-3}$.
\begin{figure*}[t]
\includegraphics[width=0.90\textwidth]{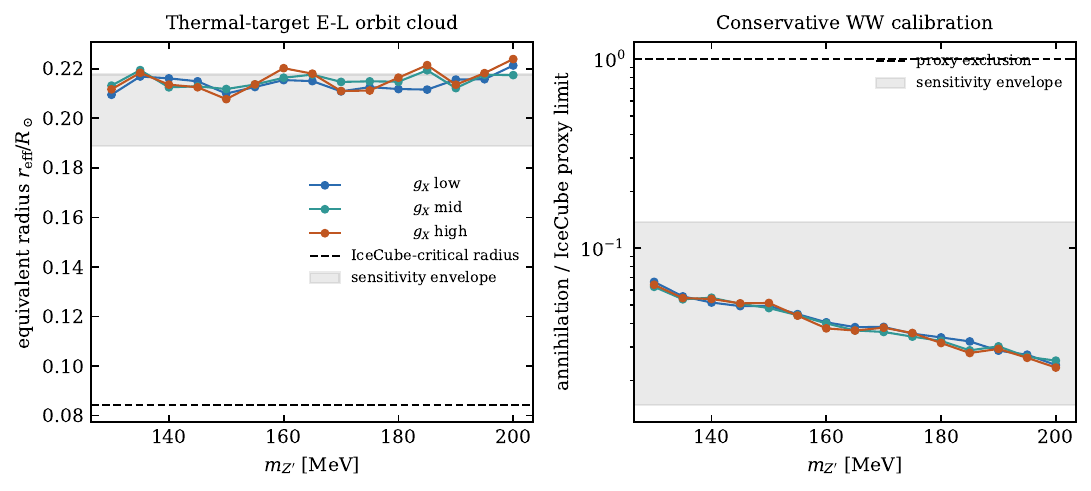}
\caption{Solar-orbit checks at 46 representative points.
Left: equivalent radius for the low-, middle-, and high-$g_X$ points
in each mediator-mass slice; the gray band shows the sensitivity
envelope. Right: corresponding annihilation rate relative to the
conservative $WW$-calibrated IceCube proxy. The horizontal dashed
lines indicate the fixed-shape reference radius and proxy boundary.
The gray band is not a statistical confidence interval.}
\label{fig:solarband}
\end{figure*}

\paragraph*{Lifetime and elastic-rate sensitivity.}
For the final benchmark, $\tau_2=137.20$ days and the scalar elastic nucleon
cross section is $0.108$ of the original transport value. Table~\ref{tab:transport_sensitivity}
starts from $\tau_{2,0}=136.5712$ days and
$\sigma_{p,0}^{\rm SI}=8.47954\times10^{-54}\,\mathrm{cm^2}$, varying either
the lifetime or both elastic nucleon cross sections by one order of magnitude.
In the elastic runs the proton and neutron amplitudes are rescaled together,
with their relative sign and recoil dependence unchanged. The remaining
transport inputs are fixed, and the lifetime runs use the same decay
kinematics.
\begin{table*}[tb]
\caption{Independent transport sensitivity tests. Each row has 1024
histories; $f_\tau=\tau_2/\tau_{2,0}$ and
$f_{\rm el}=\sigma_{\rm el}/\sigma_{{\rm el},0}$.
$R_{\rm shape}$ and $R_0$ denote the fixed-shape and no-annihilation-feedback
rate estimates divided by the same public $WW$ proxy.}
\label{tab:transport_sensitivity}
\begin{ruledtabular}
\begin{tabular}{lccccc}
Case & $f_\tau$ & $f_{\rm el}$ & $r_{\rm eff}/\rsun$ & $R_{\rm shape}$ & $R_0$\\
\hline
Central & 1 & 1 & 0.21442 & 0.04083 & 0.04087\\
$\tau_2/10$ & 0.1 & 1 & 0.21457 & 0.04098 & 0.04103\\
$10\tau_2$ & 10 & 1 & 0.21367 & 0.04109 & 0.04114\\
$\sigma_{\rm el}/10$ & 1 & 0.1 & 0.21469 & 0.04075 & 0.04080\\
$10\sigma_{\rm el}$ & 1 & 10 & 0.21070 & 0.04311 & 0.04316\\
\end{tabular}
\end{ruledtabular}
\end{table*}

Elastic initial capture is below
$4.5\times10^{-9}$ of the inelastic contribution even for the largest
elastic cross section considered, so it is omitted from the initial-state
sampling. The captured population remains spatially extended throughout the
range considered.

The relic density and other correlated observables are not recalculated in
these transport runs. The central row uses the scan capture normalization and
an independent orbit sample, giving $R_{\rm shape}=0.04083$ rather than
$0.04037$. The factors $0.1$ and $10$ are fixed sensitivity choices.

An isotope-dependent higher-order elastic prescription gives
$r_{\rm eff}=0.2093R_\odot$ and a fixed-shape $WW$ ratio of $0.0441$. This
run is used only to gauge the size of higher-order elastic effects; a complete
higher-order matching is not performed.

\paragraph*{Independent solar-profile check.}
For an independent solar-model check, we use the public ICE-CSIC
2023/SF-III GS98 and AGSS09 profiles~\cite{ICE2023}, taking the density,
temperature, mass coordinate, and elemental abundances from the tabulated
models.

Relative to the micrOMEGAs baseline, the capture rate is reduced to
$0.74036$--$0.74363$ for GS98 and $0.67893$--$0.68097$ for AGSS09.
At the benchmark point, this corresponds to
$C_\odot=6.77159\times10^{19}\,\mathrm{s^{-1}}$ and
$6.20724\times10^{19}\,\mathrm{s^{-1}}$, respectively, compared with
$9.13330\times10^{19}\,\mathrm{s^{-1}}$ for the baseline profile.

The orbital runs give $r_{\rm eff}/\rsun=0.21694$--$0.22198$. Fixed-shape
$WW$ ratios lie between $0.01710$ and $0.03273$; the largest individual value
is $0.03301$. The solar model changes the capture normalization but not the
extended character of the captured population.

\begin{figure*}[t]
\centering
\includegraphics[width=0.86\textwidth]{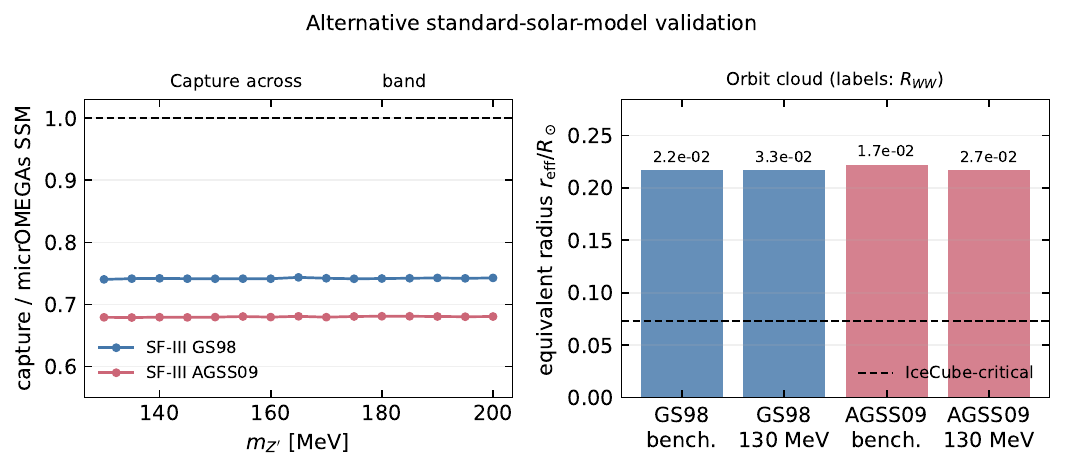}
\caption{Alternative standard-solar-model validation.  Left: capture rates
for SF-III/GS98 and SF-III/AGSS09 relative to the legacy micrOMEGAs profile on
all 15 mediator-mass slices.  Right: equivalent radii for the benchmark and the point with the
largest sampled fixed-shape proxy at $m_{Z'}=130~\mathrm{MeV}$;
labels give the corresponding $WW$-calibrated IceCube proxy ratios.}
\label{fig:altsolar}
\end{figure*}

\subsection{Relic density and theoretical consistency checks}
For the benchmark, micrOMEGAs 7.1.1 gives
\begin{align}
\Omega h^2 &= 0.11999834, \\
\langle\sigma v\rangle_0
&=4.45787\times10^{-26}\,\mathrm{cm^3\,s^{-1}} .
\end{align}
The scalar potential satisfies the tree-level copositivity conditions, and
the desired $(v,0,v_\Xi)$ vacuum is found to be the global minimum in the
numerical field scan. The Sommerfeld Born parameter is
$4.20\times10^{-3}$, while the quartic self-interaction gives
$\sigma/m=2.06\times10^{-17}\,\mathrm{cm^2\,g^{-1}}$.

For indirect detection, setting $f_{\rm eff}=1$ leaves the predicted energy
injection a factor of $6.03$ below the Planck limit. The full two-body and
cascade spectra in micrOMEGAs give a Planck ratio of $0.0257115$. The
approximate Draco, Sculptor, Leo II, and Ursa Minor analysis gives
signal-to-limit ratios of $1.9345\times10^{-4}$ for
$\Delta(-2\ln{\cal L})=3.84$ and $2.3353\times10^{-4}$ for
$\Delta(-2\ln{\cal L})=2.71$. The dwarf-galaxy numbers come from an interpolated four-target analysis
rather than the full LAT likelihood.

The elastic nucleon cross sections are
\begin{align}
\sigma_p^{\rm SI} &= 9.15965\times10^{-55}\,\mathrm{cm^2},\\
\sigma_n^{\rm SI} &= 9.34254\times10^{-55}\,\mathrm{cm^2},
\end{align}
The corresponding elastic LZ ratio is approximately $5.0\times10^{-8}$,
so the high-energy recoil signal has no appreciable elastic counterpart.

\end{document}